\documentstyle[12pt]{article}
\begin{document}

\def\ds{\displaystyle}
\def\beq{\begin{equation}}
\def\eeq{\end{equation}}
\def\bea{\begin{eqnarray}}
\def\eea{\end{eqnarray}}
\def\ve{\vert}
\def\vel{\left|}
\def\ver{\right|}
\def\nnb{\nonumber}
\def\ga{\left(}
\def\dr{\right)}
\def\aga{\left\{}
\def\adr{\right\}}
\def\lla{\left<}
\def\rra{\right>}
\def\rar{\rightarrow}
\def\nnb{\nonumber}
\def\la{\langle}
\def\ra{\rangle}
\def\ba{\begin{array}}
\def\ea{\end{array}}
\def\tr{\mbox{Tr}}
\def\ssp{{\Sigma^{*+}}}
\def\sso{{\Sigma^{*0}}}
\def\ssm{{\Sigma^{*-}}}
\def\xis0{{\Xi^{*0}}}
\def\xism{{\Xi^{*-}}}
\def\qs{\la \bar s s \ra}
\def\qu{\la \bar u u \ra}
\def\qd{\la \bar d d \ra}
\def\qq{\la \bar q q \ra}
\def\gGgG{\la g^2 G^2 \ra}
\def\q{\gamma_5 \not\!q}
\def\x{\gamma_5 \not\!x}
\def\g5{\gamma_5}
\def\sb{S_Q^{cf}}
\def\sd{S_d^{be}}
\def\su{S_u^{ad}}
\def\ss{S_s^{??}}
\def\sbp{{S}_Q^{'cf}}
\def\sdp{{S}_d^{'be}}
\def\sup{{S}_u^{'ad}}
\def\ssp{{S}_s^{'??}}
\def\sig{\sigma_{\mu \nu} \gamma_5 p^\mu q^\nu}
\def\fo{f_0(\frac{s_0}{M^2})}
\def\ffi{f_1(\frac{s_0}{M^2})}
\def\fii{f_2(\frac{s_0}{M^2})}
\def\O{{\cal O}}
\def\sl{{\Sigma^0 \Lambda}}
\def\es{\!\!\! &=& \!\!\!}
\def\ar{&+& \!\!\!}
\def\ek{&-& \!\!\!}
\def\cp{&\times& \!\!\!}
\def\se{\!\!\! &\simeq& \!\!\!}
\title{
         {\Large
                 {\bf
The $\gamma \gamma \rar H^0 H^0$ process in noncommutative
quantum electrodynamics
                 }
         }
      }

\author{\vspace{1cm}\\
{\small T. M. Aliev$^a$ \thanks
{e--mail: taliev@metu.edu.tr}\,\,,
O. \"{O}zcan$^b$ \thanks
{e-mail: oozcan@MIT.EDU}\,\,,
M. Savc{\i}$^a$ \thanks
{e--mail: savci@metu.edu.tr}} \\
{\small a Physics Department, Middle East Technical University, 
06531 Ankara, Turkey}\\
{\small b  Massachusetts Institute of Technology, Physics Department,
Cambridge, USA} }
\date{}

\begin{titlepage}
\maketitle
\thispagestyle{empty}

\begin{abstract}

We study the possibility of detecting noncommutative QED through neutral 
Higgs boson pair production at $\gamma\gamma$ collider. This is based on 
the assumption that $H^0$ interacts directly with photon as suggested 
by symmetry considerations. The sensitivity of the cross--section to 
the noncommutative scale $\Lambda_{NC}$ and Higgs mass is investigated. 
\end{abstract}

~~~PACS number(s): 11.15.-q, 11.25.Mj, 13.40.-f
\end{titlepage}

\section*{Introduction}
Noncommutative (NC) quantum field theories (NCQFT) have recently received a great
interest due to their connection to the string theories \cite{R5101}. NCQFT
provides an alternative to the ordinary quantum filed
theory, which may shed light on the study of the structure of space--time.
The main idea of NCQFT is that, in the NC space the usual
space--time coordinates $x$ are represented by operators $\hat{x}$ which
satisfy the following commutation relation
\bea
\label{e1}
\left[ \hat{x}_\mu,\hat{x}_\nu \right] = i \theta_{\mu\nu} =
\frac{i}{\Lambda_{NC}^2} C_{\mu\nu}~,
\eea
where $\Lambda_{NC}$ is the scale where NC effects become relevant, 
$C_{\mu\nu}$ is the real antisymmetric matrix with elements of order one 
and commute with ordinary $x$. In the present work we adopt
Hewett--Petriello--Rizzo parametrization \cite{R5102} for the matrix
$C_{\mu\nu}$. One might expect the scale $\Lambda_{NC}$ to be of the order
of Planck scale. However in the large extra dimension theory $[A,B]$, where
gravity becomes strong at scales of order a $TeV$, it is possible that
NC effects could be of order a $TeV$. For this reason in the
present work we consider the possibility that $\Lambda_{NC}$ may lie not too
far above the $TeV$ scale.

The matrix $C_{\mu\nu}$ is parametrized as \cite{R5102}
\bea
C_{\mu\nu} =
\left( \begin{array}{cccc} 
 0       & C_{01}  & C_{02}  &  C_{03} \\
-C_{01}  & 0       & C_{12}  & -C_{13} \\
-C_{02}  & -C_{12} & 0       &  C_{23} \\
-C_{03}  &  C_{13} & -C_{23} &  0      \\
\end{array} \right)~, \nnb
\eea
where $\sum_i \vel C_{0i} \ver^2 = 1$. Thus the matrix elements $C_{0i}$ are
related to the NC space--time components and are defined by
the direction of the background electric field $\vec{E}$. The remaining elements
$C_{ij}$ are related to the NC space--space components and are
defined the direction of the background magnetic field $\vec{B}$.  The
matrix elements $C_{0i}$ and $C_{ij}$ are parametrized as
\bea
C_{01} \es \sin\alpha \cos\beta ~, \nnb \\
C_{02} \es \sin\alpha \sin\beta ~, \nnb \\
C_{03} \es \cos\alpha ~, \nnb \\
C_{12} \es \cos\gamma ~, \nnb \\
C_{13} \es \sin\gamma \sin\beta ~, \nnb \\
C_{23} \es - \sin\gamma \cos\beta ~, \nnb
\eea
where $\beta$ defines the origin of the $\phi$ axis which we set to
$\beta=\pi/2$ and $\alpha$ and $\gamma$ are the angles of the background 
electric and magnetic fields relative to the $z$--axis.

The simplest way to construct the NCQFT from its ordinary version is by
replacing the usual product of fields in the action with the $\ast$--product of
fields
\bea
\label{e2}    
\ga f \ast g \dr (x) = \left. exp \ga \frac{i}{2} \theta^{\mu\nu}
\partial_\mu^x \partial_\nu^y \dr f(x) g(y) \ver_{x=y}~.
\eea

Noncommutative quantum electrodynamics (NCQED) based on $U(1)$ group, 
has been studied in \cite{R5103}--\cite{R5105}. Its Lagrangian is given as
\bea
\label{e3}
{\cal L} = - \frac{1}{4} F_{\mu\nu}\ast F^{\mu\nu}+
\bar{\psi}\ast (i {\cal\not\!\! D} - m ) \psi~,
\eea
where $F_{\mu\nu} = \partial_\mu A_\nu - \partial_\nu A_\mu + i e
[A_\mu,A_\nu]_\ast$ and ${\cal D}_\mu \psi = \partial_\mu\psi +
i e A_\mu \ast \psi$. Here, a generalized commutator known as the Moyal
bracket is defined as
\bea
[f,g]_\ast = f\ast g - g\ast f~.
\eea
It follows from the definition of $F_{\mu\nu}$ that, similar to the
nonabelian gauge theories, there appear both 3--point and 4--point photon
vertices resulting from the Moyal bracket term. It should be noted here
that, NC Yang--Mills theory has been studied in \cite{R5106} and
NC standard model in \cite{R5107}.

NCQFT has rich phenomenological implications due to the appearance of new
interactions. Phenomenologically the NC scale $\Lambda_{NC}$ can
take any value. However, recent studies in extra dimensions show that
gravity becomes strong at the $TeV$ scale \cite{R5108,R5109}. So it is
possible that NC effects could set in at a $TeV$. Therefore we consider the
case when $\Lambda_{NC}$ is not too far above the $TeV$ scale.

A series of phenomenological studies of NCQED at next--generation 
high energy linear collider have already been carried out in
\cite{R5110} and in \cite{R5111,R5112}. Also, the fermion and charged 
Higgs boson production at $\gamma \gamma$ collider has been studied in 
\cite{R5111}. The feasibility of detecting NCQED through neutral Higgs boson
pair production at linear colliders, assuming that $H^0$ interacts directly
with photon, has been considered in \cite{R5113}. 

The next--generation linear colliders (NLC) are planned to operate in 
$e^+ e^-,~\gamma\gamma$ and $\gamma e$ modes. It is well known that, at 
high energy and luminosity, $e^+ e^-$ collider can be converted into
$\gamma\gamma$ collider, practically with almost the same energy and
luminosity, using the laser backscattering technique \cite{R5114}.

In the present work we consider the possibility of testing the NC effects at
NLC in the $\gamma\gamma$ mode by studying the 
$\gamma\gamma \rar H^0 H^0$ process.

We begin our calculation, following
\cite{R5113}, by assuming that the neutral particle also participates in the
electromagnetic interaction, i.e.,
\bea
\label{e5}
{\cal L}_H = \frac{1}{2} \ga {\cal D}_\mu H^0 \ast {\cal D}^\mu H^0 \dr~,
\eea
where
\bea
{\cal D}_\mu H^0 = \partial_\mu H^0 + i e [A_\mu,H^0]_\ast~.\nnb
\eea
A direct result of the interactions in Eqs. (\ref{e3}) and (\ref{e5}) leads
to the relevant Feynman rules which are presented in Fig. (1) and the
related Feynman diagrams are shown in Fig. (2). It follows from these Feynman 
rules that when $\theta_{\mu\nu} \rar 0$, which corresponds to ordinary 
quantum electrodynamics (QED), all interaction vertices go to zero. In other
words, this process is forbidden at tree--level in ordinary QED. Therefore
contribution to this channel comes completely from NCQED and hence this
process can serve as a good possibility of testing the grounds of NCQED.

The amplitude for the $\gamma\gamma \rar H^0 H^0$ process can be written in
the following form
\bea
\label{e6}
{\cal M} \es - 4 i e^2 \varepsilon_\alpha (k_1) \varepsilon_\beta (k_2) 
\Bigg\{ 4 p_{1\alpha} p_{2\beta} \frac{1}{\hat{t}-m_H^2} 
\sin(k_1 \frac{\theta}{2} p_1) \sin(k_2 \frac{\theta}{2} p_2) \nnb \\
\ar 4 p_{1\beta} p_{2\alpha} \frac{1}{\hat{u}-m_H^2} 
\sin(k_2 \frac{\theta}{2} p_1) \sin(k_1 \frac{\theta}{2} p_2) \nnb \\
\ar  \frac{1}{\hat{s}} \left[(k_1-k_2)(p_1-p_2) g_{\alpha\beta} +
2 k_{2\alpha}(p_1-p_2)_\beta - 2 k_{2\beta} (p_1-p_2)_\alpha \right] 
\sin(k_1 \frac{\theta}{2} k_2) \sin(p_1 \frac{\theta}{2} p_2) \nnb \\
\ar g_{\alpha\beta} 
\left[ \sin(k_1 \frac{\theta}{2} p_1) \sin(k_2 \frac{\theta}{2} p_2) +
\sin(k_1 \frac{\theta}{2} p_2) \sin(k_2 \frac{\theta}{2} p_1) \right]
\Bigg\}~,
\eea 
where $\varepsilon_\alpha (k_1)$ and $\varepsilon_\beta (k_2)$ are the
photon polarization vectors, $p_1$ and $p_2$ are the Higgs boson momenta,
respectively, and $\hat{s},~\hat{t}$ and $\hat{u}$ are the usual Mandelstam
variables.

At this point we would like to make the following remark. Due to the
presence of the triple photon vertex, computation of
$\vel {\cal M} \ver^2$ and summing over photon polarization must be handled
carefully to make sure that the Ward identities are satisfied and to
guarantee that the unphysical photon polarization states do not appear. For
this aim we will follow to different approaches to the present problem.
In the first method one can use explicit transverse photon polarization
vectors. The second method could be that one can use the physical
polarization sums for the photons so that only the physical polarization
contributes to $\vel {\cal M} \ver^2$. A convenient form is
\bea
\label{e7}
\sum_\lambda \varepsilon^\mu \varepsilon^{\ast\nu} (\lambda) = 
- \left[ g^{\mu\nu} - 
\frac{n^\mu k^\nu + n^\nu k^\mu}{n k} +
\frac{n^2 k^\mu k^\nu}{(n k)^2}\right]~,
\eea
where $n$ is any arbitrary vector. In further analysis we will set $n^2=0$
and $k_1 n_i \neq 0$, which corresponds to the axial gauge. In practice it
is most convenient to choose $n_i$ as the photon momentum.

The unpolarized differential cross--section in the $\gamma\gamma$ center of mass is
given by

\bea
\frac{d^2\hat{\sigma}}{dz d\varphi} = \frac{\hat{v}\alpha^2}{2 \hat{s}} \vel
{\cal M} \ver^2~,\nnb
\eea
where
\bea
\label{e8}
\lefteqn{
\vel {\cal M} \ver^2 = 
\frac{1}{\hat{s}^2 (m_H^2 -\hat{t})^2 (m_H^2 -\hat{u})^2} \Bigg\{
-4 \sin (k_1 \frac{\theta}{2} k_2) 
   \sin (k_1 \frac{\theta}{2} p_1)
   \sin (k_2 \frac{\theta}{2} p_2)
   \sin (p_1 \frac{\theta}{2} p_2)} \nnb \\
\cp (m_H^2 - \hat{t}) (m_H^2 - \hat{u})^2 (\hat{t} - \hat{u})
\Big\{ 
       2 m_H^4 - \hat{s}^2 - \hat{s} \hat{t} + \hat{t}^2 +
       \hat{u}^2 + m_H^2 \Big[3 \hat{s}- 2 (\hat{t} + \hat{u})\Big] 
\Big\} \nnb \\
\ar
4  \sin (k_1 \frac{\theta}{2} p_1)
   \sin (k_1 \frac{\theta}{2} p_2)
   \sin (k_2 \frac{\theta}{2} p_1)
   \sin (k_2 \frac{\theta}{2} p_2)
(m_H^2 - \hat{t}) (m_H^2 - \hat{u}) \nnb \\
\cp \Big\{
      8 m_H^8 + 2 \hat{s}^4 + \hat{s}^3 (\hat{t} + \hat{u}) +
\hat{s}^2 (-4 \hat{t}^2 + \hat{t} \hat{u} - 4 \hat{u}^2 )
+ 2 (\hat{t}^2+\hat{u}^2)^2 -
\hat{s} (\hat{t}^3 + \hat{t}^2 \hat{u} + \hat{t} \hat{u}^2 + \hat{u}^3) \nnb \\
\ar
4 m_H^6 \Big[ 5 \hat{s} - 4 (\hat{t} + \hat{u}) \Big] +
m_H^4 \Big[ 5 \hat{s}^2 - 22 \hat{s} (\hat{t} + \hat{u}) + 
16 (\hat{t}^2 + \hat{t} \hat{u} + \hat{u}^2) \Big] \nnb \\
\ar
m_H^2 \Big[ -10 \hat{s}^3 + 5 \hat{s}^2 (\hat{t} + \hat{u}) +
4 \hat{s} (3 \hat{t}^2 + \hat{t} \hat{u} + 3 \hat{u}^2 ) -
8 (\hat{t}^3 + \hat{t}^2 \hat{u} + \hat{t} \hat{u}^2 + \hat{u}^3)\Big]
\Big\} \nnb \\
\ek
4 \sin (k_1 \frac{\theta}{2} k_2) 
   \sin (k_1 \frac{\theta}{2} p_2)
   \sin (k_2 \frac{\theta}{2} p_1)
   \sin (p_1 \frac{\theta}{2} p_2)
(m_H^2 - \hat{t})^2 (m_H^2 - \hat{u}) (\hat{t} - \hat{u}) \nnb \\
\cp
\Big\{ 
       2 m_H^4 - \hat{s}^2 + \hat{t}^2 - \hat{s} \hat{u} +
       \hat{u}^2 + m_H^2 \Big[3 \hat{s}- 2 (\hat{t} + \hat{u})\Big] 
\Big\} \nnb \\
\ar
2 \sin^2(k_1 \frac{\theta}{2} k_2) 
  \sin^2(p_1 \frac{\theta}{2} p_2)
(m_H^2 - \hat{t})^2 (m_H^2 - \hat{u})^2 (\hat{t} - \hat{u})^2 \nnb \\
\ar
2 \sin^2(k_1 \frac{\theta}{2} p_1)
  \sin^2(k_2 \frac{\theta}{2} p_2)
  (m_H^2 - \hat{u})^2 \nnb \\
\cp 
\Big\{
       8 m_H^8 - 4 m_H^6 \Big[3 \hat{s} + 4 (\hat{t} + \hat{u})\Big] +
\hat{t} \Big[ 2 \hat{s}^3 + \hat{s}^2 \hat{t} + 8 \hat{t} \hat{u}^2 -
2 \hat{s} (\hat{t}^2 + \hat{u}^2)\Big] \nnb \\
\ar
m_H^4 \Big[13 \hat{s}^2 + 4 \hat{s} (2 \hat{t}+3 \hat{u}) +
8 (\hat{t}^2 + 4 \hat{t} \hat{u} + \hat{u}^2) \Big] - 
2 m_H^2 \Big[\hat{s}^3 + 3 \hat{s}^2 \hat{t} + 
8 \hat{t} \hat{u} (\hat{t}+\hat{u}) \nnb \\
\ek \hat{s}(3 \hat{t}^2 - 6 \hat{t} \hat{u}+ \hat{u}^2) \Big] 
\Big\}
+
2 \sin^2(k_1 \frac{\theta}{2} p_2)
  \sin^2(k_2 \frac{\theta}{2} p_1)
  (m_H^2 - \hat{t})^2 
\Big\{
       8 m_H^8 - 4 m_H^6 \Big[3 \hat{s} + 4 (\hat{t} + \hat{u})\Big] \nnb \\
\ar
\hat{u} \Big[ 2 \hat{s}^3 + \hat{s}^2 \hat{u} + 8 \hat{t}^2 \hat{u} -
2 \hat{s} (\hat{t}^2 + \hat{u}^2)\Big] +
m_H^4 \Big[13 \hat{s}^2 + 4 \hat{s} (3 \hat{t}+2 \hat{u}) +
8 (\hat{t}^2 + 4 \hat{t} \hat{u} + \hat{u}^2) \Big] \nnb \\
\ek
2 m_H^2 \Big[\hat{s}^3 + 3 \hat{s}^2 \hat{u} + 
8 \hat{t} \hat{u} (\hat{t}+\hat{u})-
\hat{s}(\hat{t}^2 - 6 \hat{t} \hat{u}+ 3\hat{u}^2) \Big]
\Big\}
\Bigg\}~.
\eea
Here,
\bea
\hat{s} \es (k_1+k_2)^2=(p_1+p_2)^2~,\nnb \\
\hat{t} \es (k_1-p_1)^2=(k_2-p_2)^2=m_H^2 - \frac{\hat{s}}{2} (1 - \hat{v} z)~,\nnb \\      
\hat{u} \es (k_1-p_2)^2=(k_2-p_1)^2=m_H^2 - \frac{\hat{s}}{2} (1 + \hat{v} z)~,\nnb
\eea
where $\hat{v} = \sqrt{1 - 4 m_H^2/\hat{s}}$ is the velocity of the Higgs
boson, $z=\cos\theta$ and $\theta$ is the angle between $\vec{k}_1$ (the
$z$--direction) and $\vec{p}_1$ three--momenta, and $\varphi$ is the
azimuthal angle.

Before performing numerical analysis we would make the following remark.
Firstly, it might seem that Eq. (8) could be used for numerical analysis.
In general, however, it is not directly applicable for a real collider
experiment and we have the following problems in interpretation of the
experimental data. The first problem is connected with the existence of two
cross--sections, which we can briefly explain as follows. As has already
been mentioned earlier, our analysis is carried out in the photon--photon center
of mass frame and not in the laboratory frame in which the center of mass
frame can be boosted. This due to the fact that the colliding photons
generally do not have equal energies. Since the theory is no longer Lorentz
invariant, these two cross-sections are no longer simply related to each
other. In principle, this may change the numerical results significantly.
There is, also, the additional issue in regard to the orientation of the
reference frame with respect to some cosmological reference frame.
This is due to the fact that $\theta_{\mu\nu}$ is not a Lorentz
tensor, and as a result, if it is defined in one reference frame, it should
change with respect to another reference frame under space--time coordinate
transformations. If we neglect the change in magnitude of $\vec{\theta}$ in
the local reference frame, the change in $\vec{\theta}$ in direction
relative to the local reference frame must be taken into account, i.e., the
earth's rotation needs to be taken into account in the analysis of the
experimental data. The rotation of earth leads to the following
distributions:
\begin{itemize}
\item distribution over local $\theta$ and $\varphi$ angles when averaging
over earth's rotation is performed,
\item distribution over earth's rotation which leads to the day--night
effects.
\end{itemize}
In the present work we neglect the effects coming from earth's rotation and
we are planning to discuss these ditributions in detail in one of our
forthcoming works.          

In practice, it is very difficult to produce high--energy monochromatic photon
beams. As has already been noted, a realistic method to obtain high--energy
photon beam is to use the laser back--scattering technique on an electron or
positron beam which produces abundant hard photons nearly along the same
direction as the original electron or positron beam. However, the photon beam energy
obtained this way is not monochromatic. The energy spectrum of the
back--scattered photon is given by \cite{R5116}
\bea
\label{e9}
f(x) \es \frac{1}{D(\xi)} \left[ 1 - x + \frac{1}{1-x} -
\frac{4 x}{\xi (1-\xi)} + \frac{4 x^2}{\xi^2 (1-x)^2}\right]~,\nnb \\
D(x) \es \ga 1-\frac{4}{\xi} - \frac{8}{\xi^2}\dr \ln (1+\xi) + \frac{1}{2} +
\frac{8}{\xi} - \frac{1}{2(1+\xi)^2}~,
\eea
where $x$ is the fraction of energy of the incident $e^\pm$ beam,
$\xi = 2 (1+\sqrt{2})$ and $x_{max} = \xi/(1+\xi) \approx 0.828$

The cross--section at such a $\gamma\gamma$ collider with the $e^+ e^-$
center of mass frame energy $\sqrt{s}$ is given by 
\bea 
\label{e10}
\sigma = \int_{x_{1min}}^{x_{max}} dx_1 f(x_1) 
         \int_{x_{2min}}^{x_{max}} dx_2 f(x_2)   
\int_{-1}^{+1} dz \int_{0}^{2 \pi} d \varphi\,
\frac{d^2\sigma (x_1,x_2,s,z,\varphi)}{dz d\varphi}~,
\eea
where 
\bea
x_{1min} = \frac{4 m_H^2}{s x_{max}},~~~~\mbox{\rm and} ~~~~
x_{2min} = \frac{4 m_H^2}{s x_1}~. \nnb
\eea

In further numerical analysis we consider linear $e^+ e^-$ colliders 
operating at $\sqrt{s}=1$--$1.5~TeV$ (NLC proposal) \cite{R5117}, and 
$\sqrt{s}=3~TeV$ \cite{R5118}. As has already been
mentioned, we take $\beta=\pi/2$. Therefore, among all components of the
matrix $C_{\mu\nu}$ the ones that survive are
$C_{02},~C_{03},~C_{12}$ and $C_{13}$.  

In Figs. (3) and (4) ((5) and (6)), we present the dependence of the 
cross--section of the $\gamma \gamma \rar H^0 H^0$ process on the NC 
geometry parameter $\Lambda_{NC}$ and Higgs boson mass $m_H$ at 
$\alpha=\pi/2$ and $\alpha=0$, and at $\sqrt{s}=1.5~TeV$ (at 
$\sqrt{s}=3~TeV$), respectively. In Figs.  (7) and (8), we depict the
dependence of the cross--section on the NC 
geometry parameter $\Lambda_{NC}$ and Higgs boson mass $m_H$ at 
$\gamma=0$, and at $\sqrt{s}=1.5~TeV$ and  
$\sqrt{s}=3~TeV$, respectively. 

When all figures are taken into account, we observe that the cross--section
gets larger values only for the $C_{03}$ matrix element compared to the
other cases. This fact can be explained in the following way. The
expressions with the coefficients $C_{02}$ and $C_{13}$ have azimuthal angle
$\varphi$ dependence, while $C_{03}$ is independent of $\varphi$. In order to
calculate the cross--section we perform integration over $\varphi$. In doing
so, terms that have $\varphi$ dependence become zero, but rest of the terms
that are independent of $\varphi$ are just multiplied by $2\pi$. Obviously,
this is the reason why cross--section gets larger value for the $C_{03}$
case. Note that, stronger constraints to the parameter
$C/\Lambda_{NC}^2$, where $C$ is the value of the elements of the matrix
$C_{\mu\nu}$, were obtained in \cite{R5119}. 

Finally, we would like discuss the following issue. In the SM this process
can take place via the loop diagram. In answering the question whether the
given process takes place via the NC effects or SM loop effects, it is
better to consider the azimuthal angle dependence of the cross--section.
In the NC approach this process depends explicitly on the azimuthal
angle $\varphi$ through $k_1 \theta k_2$, while it contains no explicit
dependence on $\varphi$ if the same process takes place via the loop effects
in the SM. So, an investigation of the cross--section on the azimuthal angle
$\varphi$ can give unambiguous information about the existence of the
noncommutative geometry effects. In this connection, the dependence of the
cross--section of the considered process on $\varphi$, at two different fixed
values of $\Lambda_{NC}$ and $m_H$, and at $\sqrt{s}=3~TeV$, are presented
in Figs. (9) and (10), for the cases $C_{02}$ and $C_{03}$, respectively.

In summary, we have examined the $\gamma\gamma \rar H^0 H^0$ process, which
is strictly forbidden in the SM at tree level, in establishing
noncommutative geometry. Our analysis yields that the cross--section is more
sensitive to the matrix element $C_{03}$ and analysis of the cross--section
on the azimuthal angle is a potentially efficient tool in establishing NC
effects.

\newpage

\section*{Figure captions}
{\bf Fig. (1)} Feynman rules for the $\gamma\gamma \rar H^0 H^0$ process in
NCQED.\\ \\ 
{\bf Fig. (2)} Feynman diagrams for the $\gamma\gamma \rar H^0 H^0$ process
in NCQED.\\ \\
{\bf Fig. (3)} The dependence of the cross--section for the $\gamma\gamma
\rar H^0 H^0$ process on $\Lambda_{NC}$ and $m_H$, at $\alpha = \pi/2$ and 
 at $\sqrt{s}=1.5~TeV$.\\ \\
{\bf Fig. (4)} The dependence of the cross--section for the $\gamma\gamma   
\rar H^0 H^0$ process on $\Lambda_{NC}$ and $m_H$, at $\alpha = 0$ and 
 at $\sqrt{s} = 1.5~TeV$.\\ \\
{\bf Fig. (5)} The same as in Fig. (3), but at $\sqrt{s} = 3~TeV$.\\ \\
{\bf Fig. (6)} The same as in Fig. (4), but at $\sqrt{s} = 3~TeV$.\\ \\
{\bf Fig. (7)} The dependence of the cross--section for the $\gamma\gamma   
\rar H^0 H^0$ process on $\Lambda_{NC}$ and $m_H$, at $\gamma = \pi/2$ and 
 at $\sqrt{s}=1.5~TeV$.\\ \\
{\bf Fig. (8)} The same as in Fig. (7), but at $\sqrt{s} = 3~TeV$.\\ \\
{\bf Fig. (9)} The dependence of the cross--section for the $\gamma\gamma
\rar H^0 H^0$ process on the azimuthal angle $\varphi$, at $\alpha = \pi/2$
and $\sqrt{s} = 3~TeV$,
and at two different values of $\Lambda_{NC}=500~GeV;600~GeV$, and 
$m_H=150~GeV;300~GeV$.\\ \\
{\bf Fig. (10)} The same as in Fig. (9), but at $\alpha = 0$.
\newpage

\begin{figure}
\vskip -0.5 cm
    \includegraphics{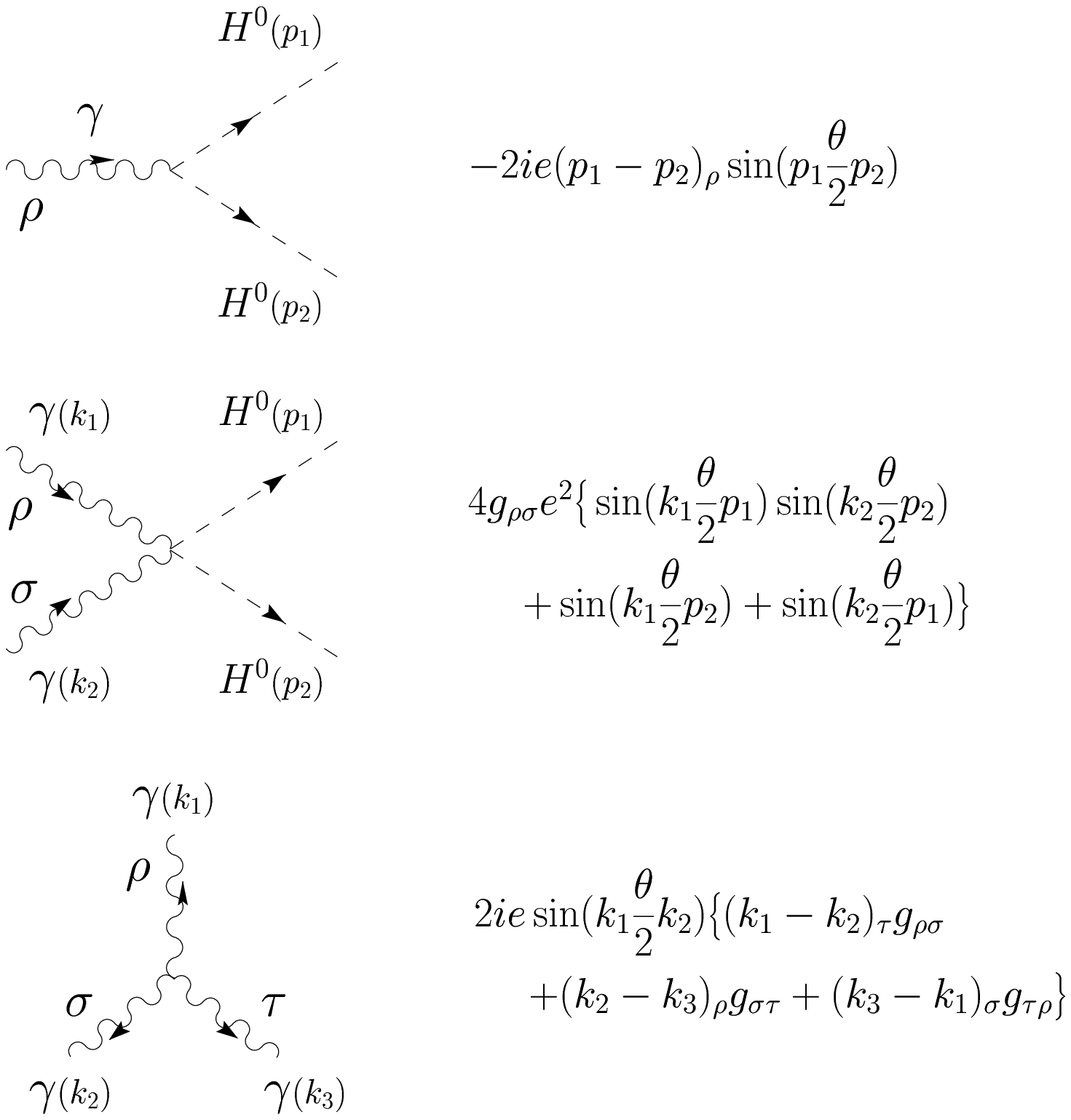}
\vskip 12.5 cm
\caption{}
\end{figure}

\begin{figure}
\vskip 1 cm
    \includegraphics{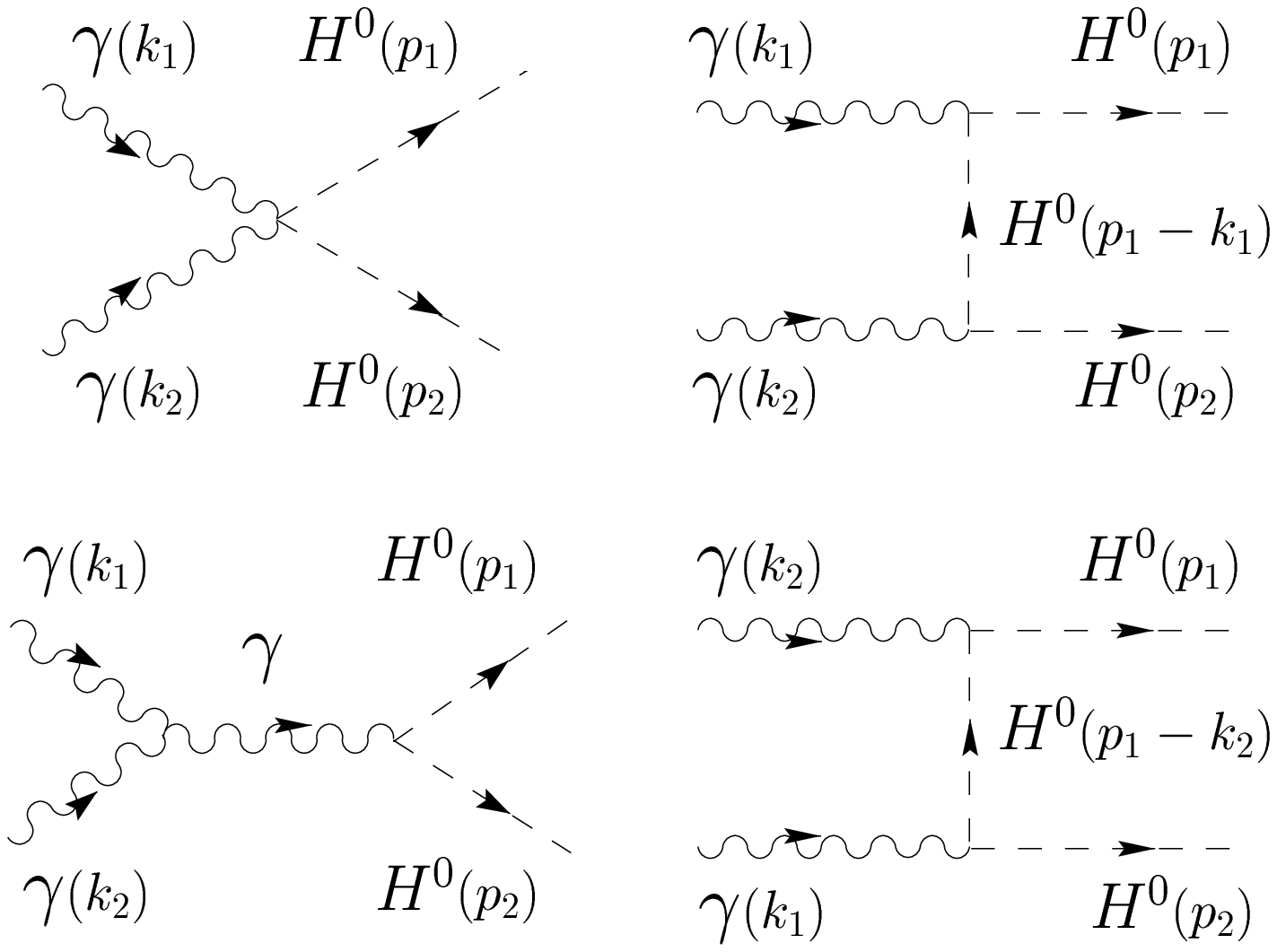}
\vskip 7.25 cm
\caption{}
\end{figure}

\begin{figure}  
\vskip 0 cm   
    \includegraphics{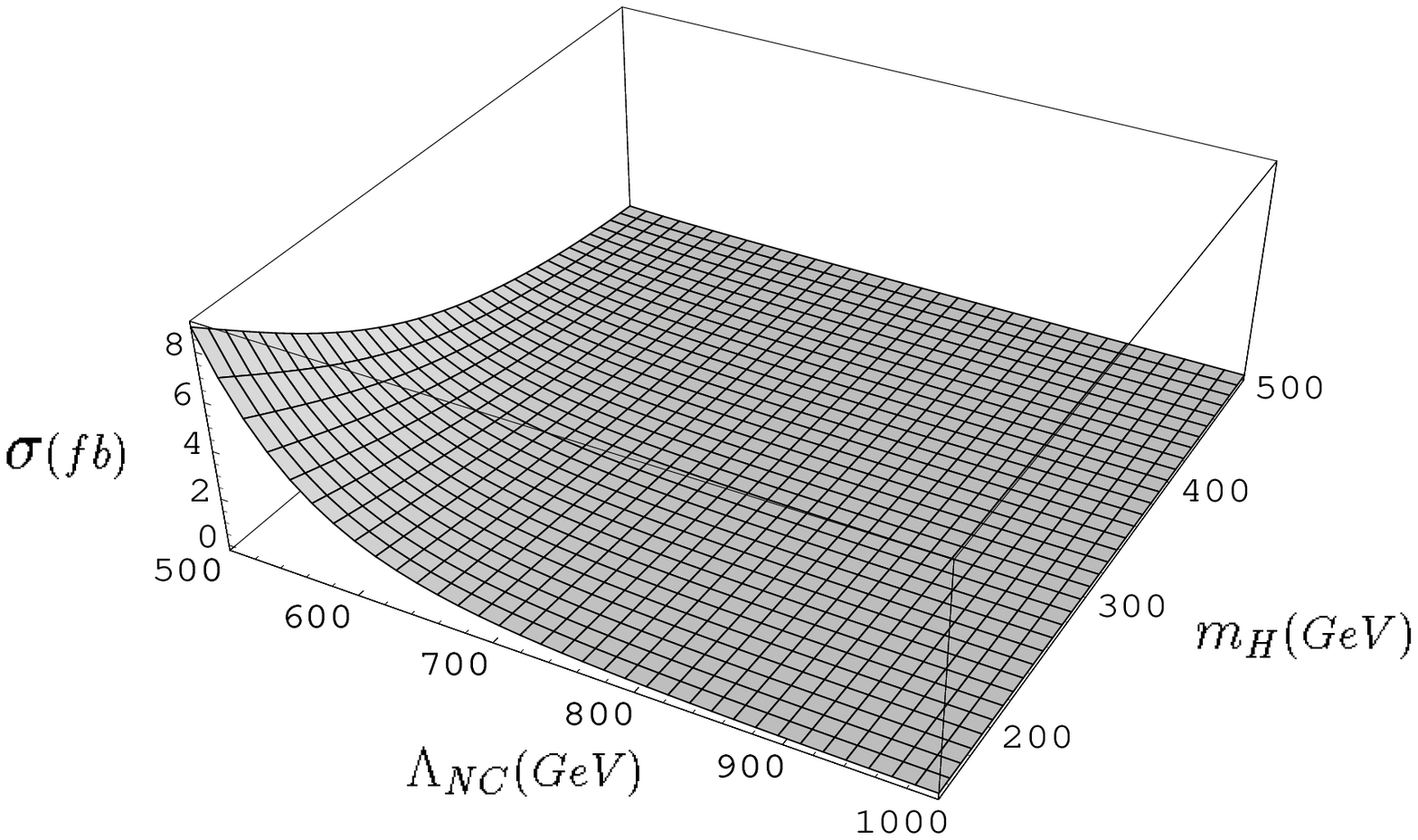}
\vskip 7.0cm     
\caption{}
\end{figure}

\begin{figure}
\vskip 1. cm
    \includegraphics{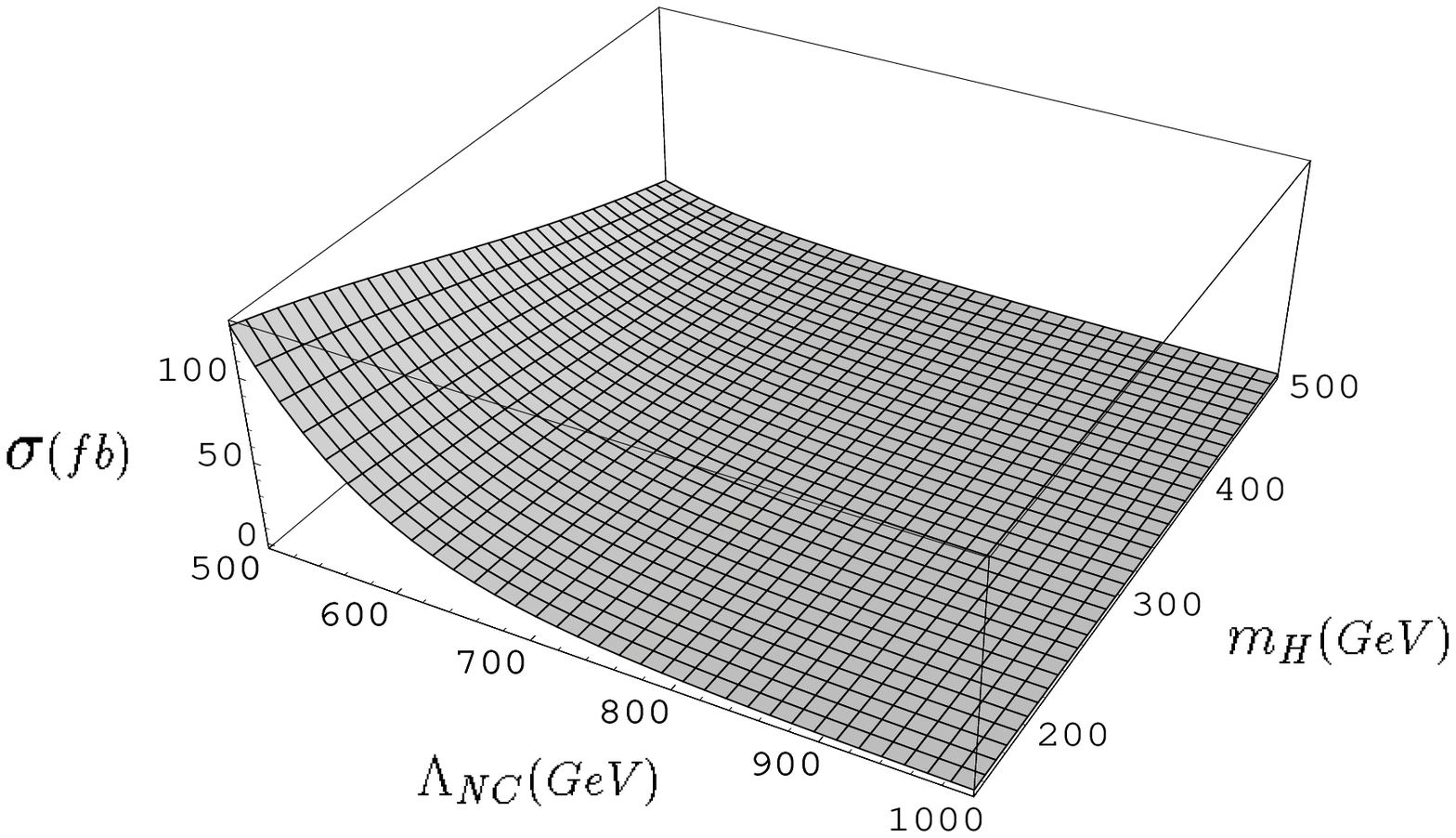}
\vskip 8.0 cm
\caption{}
\end{figure}

\begin{figure}  
\vskip 1.5 cm   
    \includegraphics{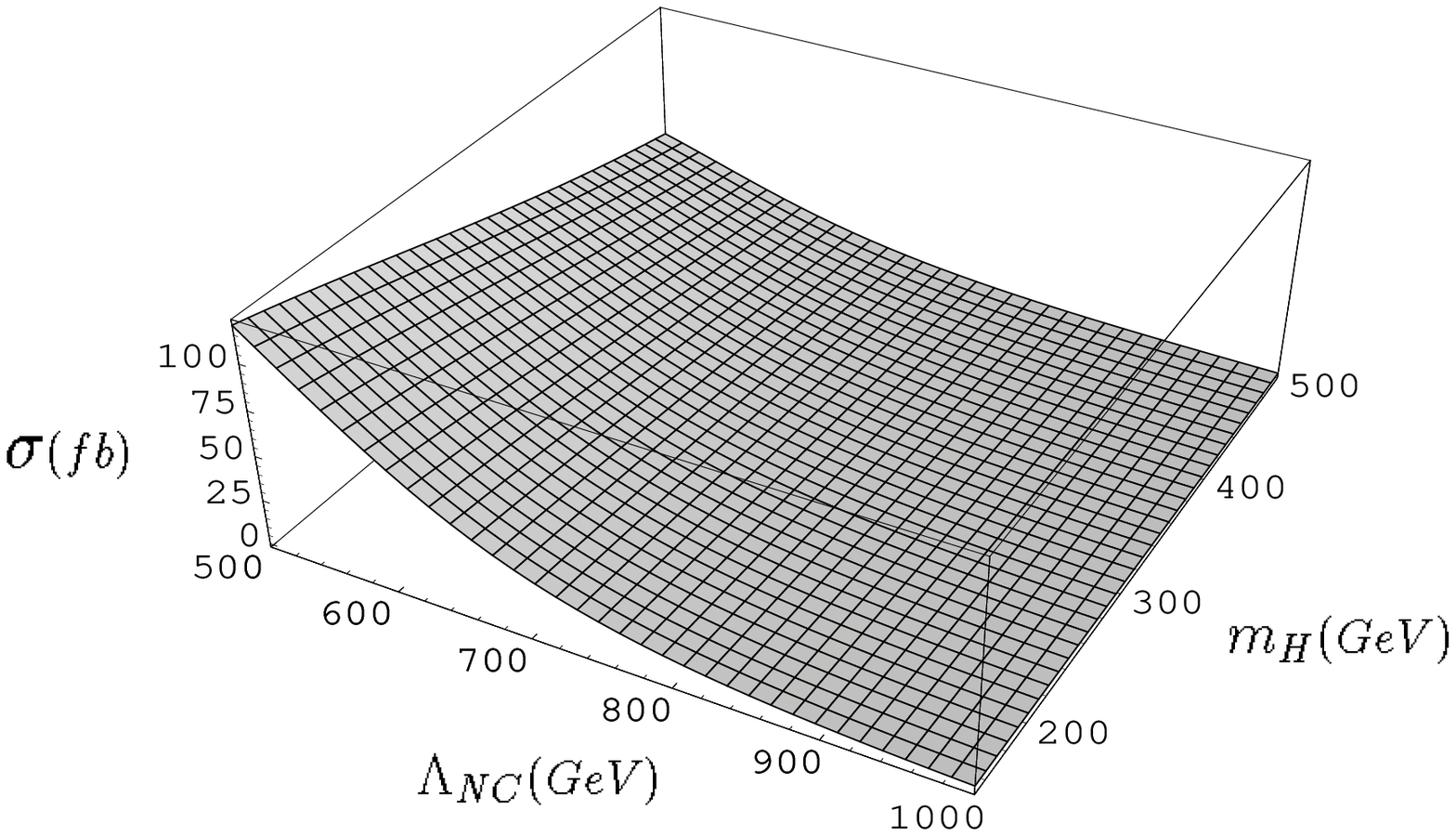}
\vskip 7.0cm     
\caption{}
\end{figure}

\begin{figure}
\vskip 1. cm
    \includegraphics{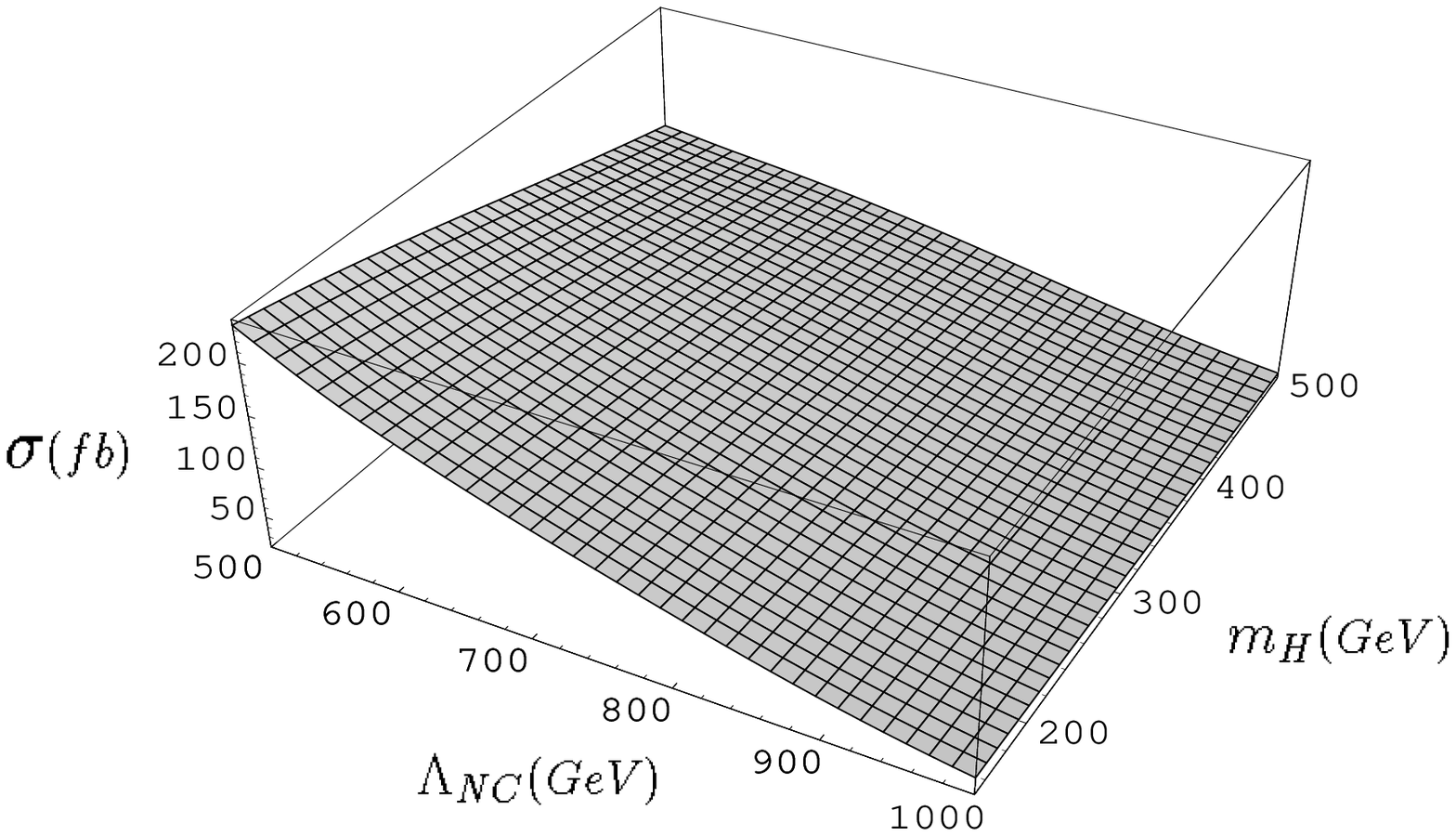}
\vskip 8.0 cm
\caption{}
\end{figure}

\begin{figure}  
\vskip 1.5 cm   
    \includegraphics{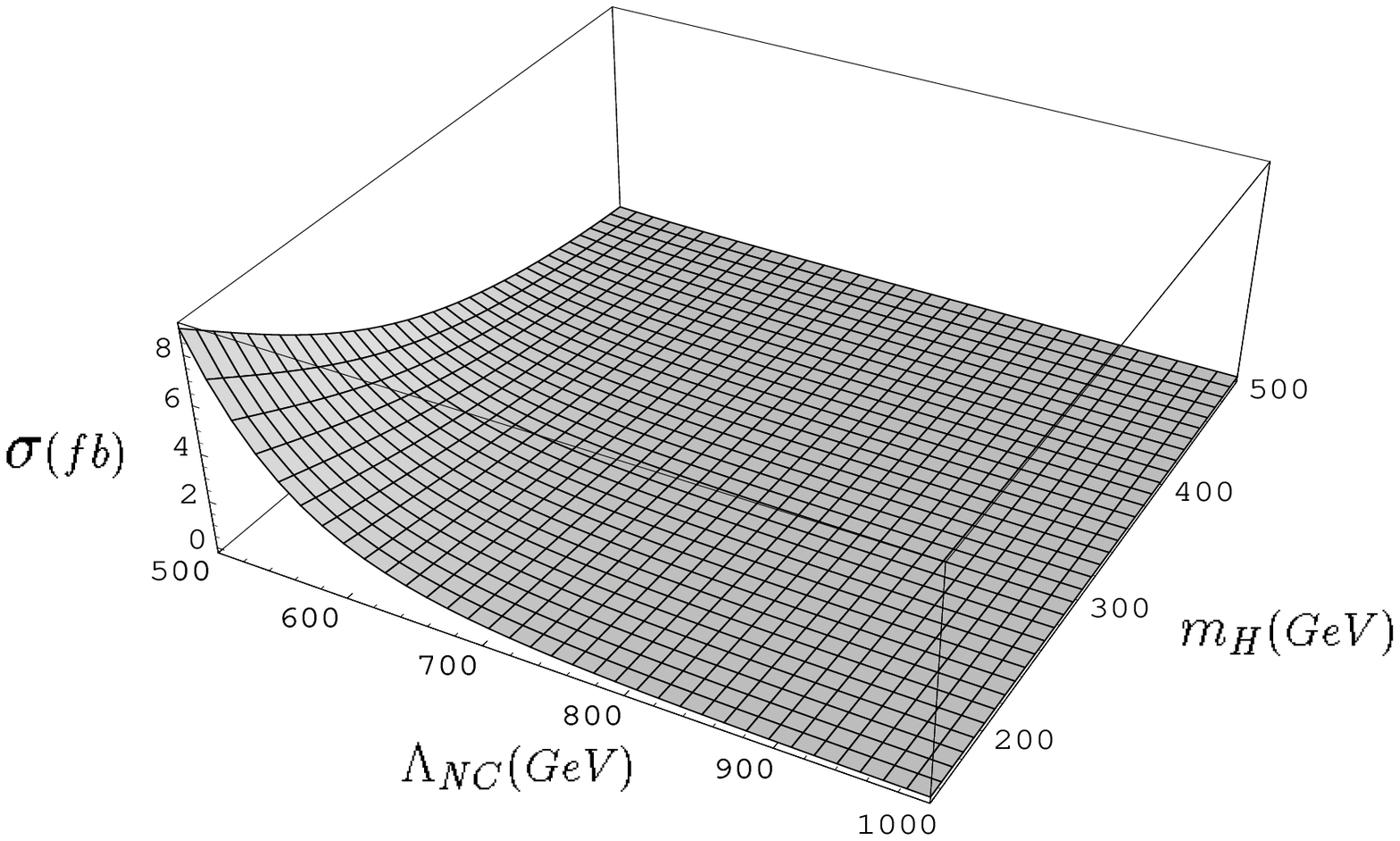}
\vskip 7.0cm     
\caption{}
\end{figure}

\begin{figure}
\vskip 1. cm
    \includegraphics{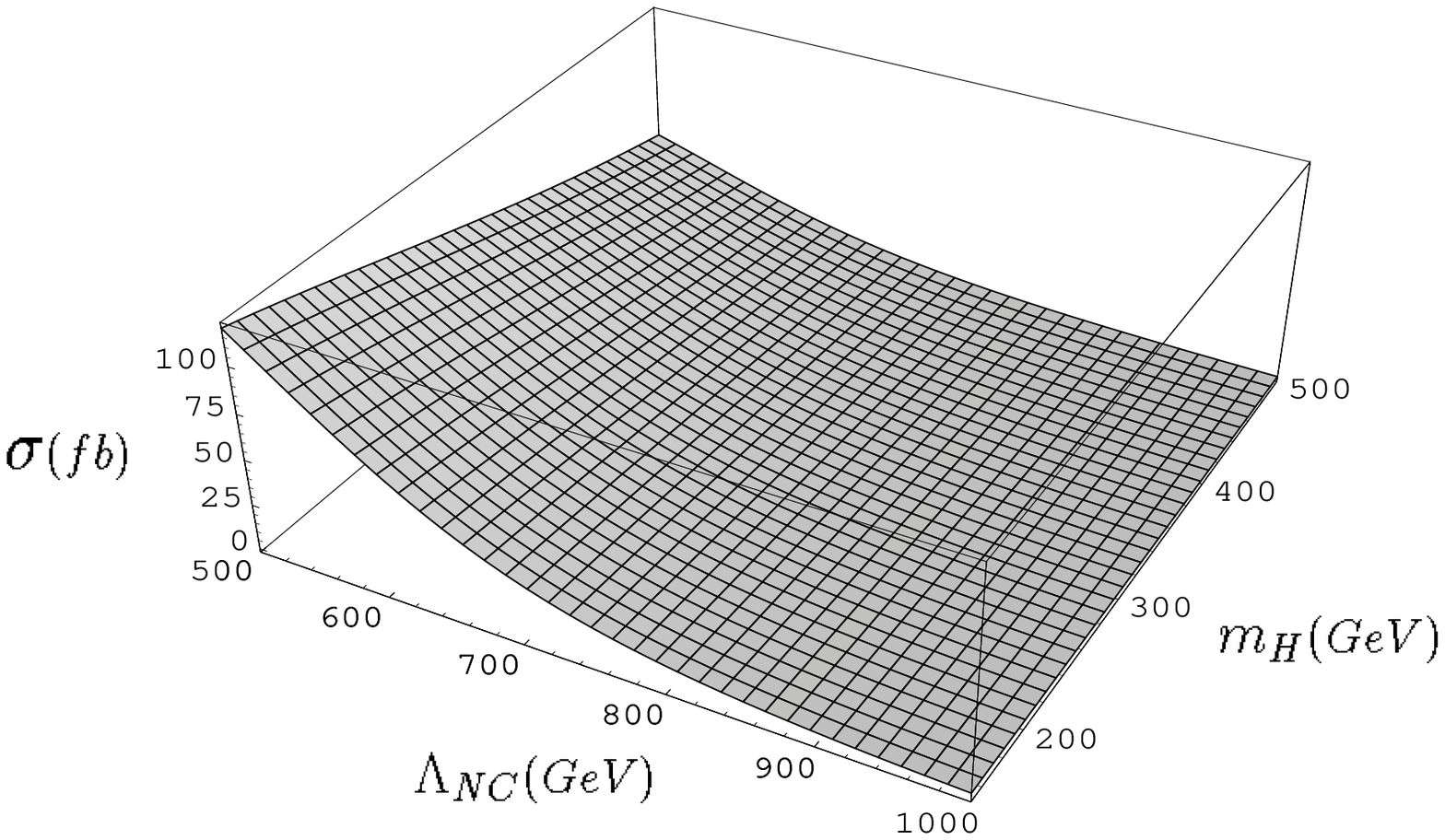}
\vskip 8.0 cm
\caption{}
\end{figure}

\begin{figure}
\vskip 1.5 cm
    \includegraphics{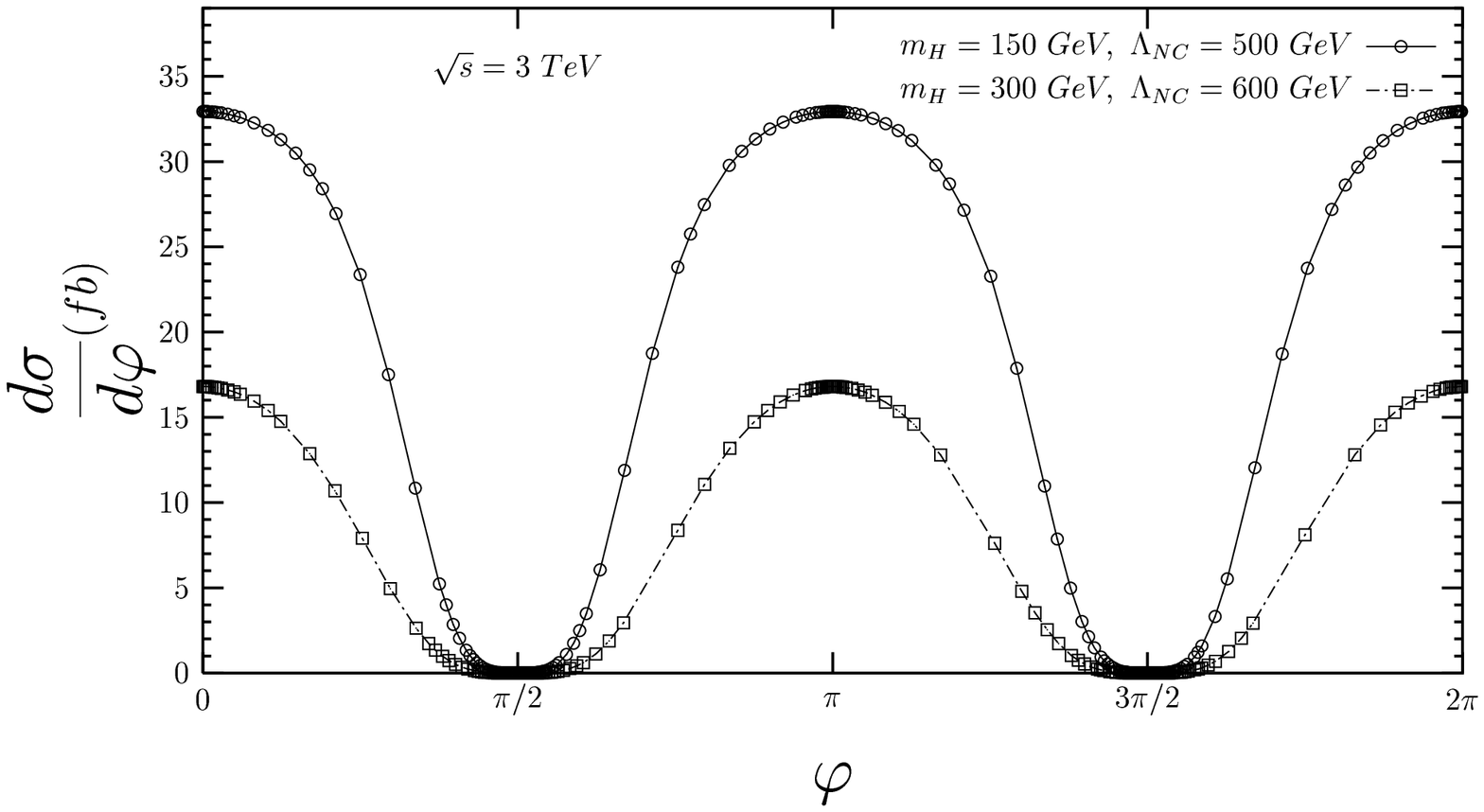}
\vskip 7.8cm
\caption{}
\end{figure}

\begin{figure}
\vskip 2.5 cm
    \includegraphics{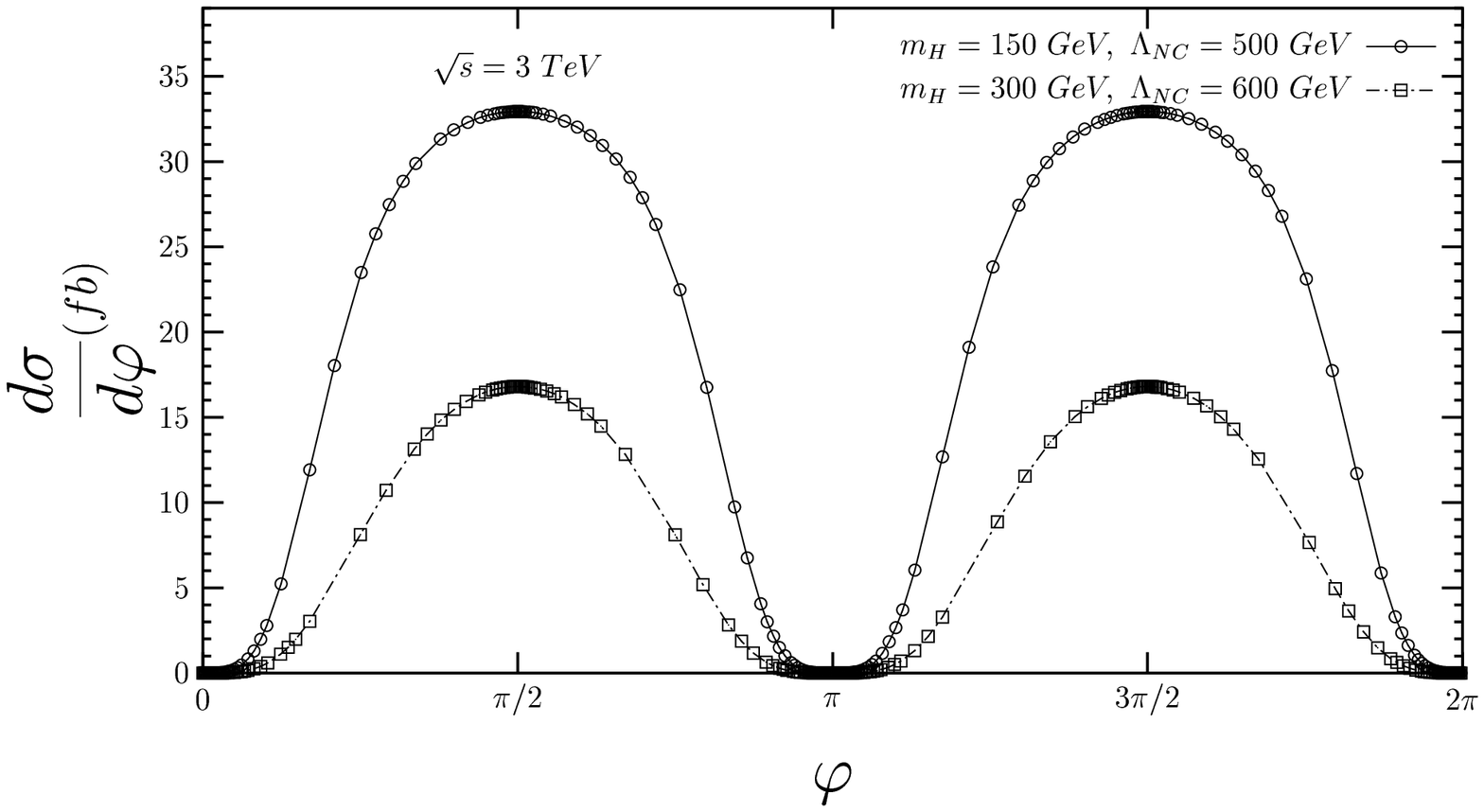}
\vskip 7.8 cm
\caption{}
\end{figure}

\end{document}